\documentclass[a4paper,12pt]{article}
\usepackage{epsfig}
\usepackage{amssymb}
\begin{document}
\baselineskip=17pt
\title{Schwinger- Dyson Equations and Dynamical Symmetry 
Breaking in Quantum R$^2$- gravity}
\author{Yu. I. Shil'nov\thanks{Supported by Deutscher Akademischer 
Austauschdienst, e-mail: shilnov@itp.uni-leipzig.de. Permanent e-mail: 
shilnov@express.kharkov.ua.  }\\
{\it Department of Theoretical Physics, Faculty of Physics,}\\
{\it Kharkov State University,}\\
{\it Svobody Sq. 4, 310077, Kharkov, Ukraine}\\
{\it and}\\
{\it Center for Advanced Studies, Leipzig University,}\\
{\it Augustusplatz 10, D- 04109, Leipzig, Germany}\\[1.5ex]
V. V. Chitov, A. T. Kotwicki\\
{\it Department of Theoretical Physics, Faculty of Physics},\\
{\it Kharkov State University},\\ 
{\it Svobody Sq. 4, 31077, Kharkov, Ukraine}}
\date{ }
\maketitle
\large

\vskip 1cm
The dymamical chiral symmetry breaking  in higher- derivative quantum 
gravity
has been investigated on the flat background. The  Schwinger- Dyson 
equations  
numerical solutions have been found in the ladder approximation. Both 
two-  and four- dimensional cases have been considered. The dymamical 
fermion mass 
generation  accompanied by the second- order phase transition has 
been shown to take place 
in a different gauges.

\vfill\eject

\section*{1. Introduction.}

The dynamical symmetry breaking (DSB) has been considered as a possible mechanism
of appropriable fermion mass generation in the quantum field theory for a 
long time [1- 20]. Usually, the broken symmetry is the chiral one, which plays 
a very important role in the high energy physics [3- 12, 18, 19]. There are some methods
of DSB investigation: the direct calculation of composite
fields effective action [1, 6, 7, 16], Schwinger- Dyson equations (SDE) analysis 
[3- 8, 13, 15, 17, 18] or
solution of composite field renormalizations group equations [20].  

These methods have been generalized for curved spacetime case  in 4D QED [21], 
four- fermionic models [22- 28], Einstein general relativity [29] and 2D induced gravity 
[30, 31]. The results of the DSB in curved spacetime have been the subject of considerable 
attention because the curvature induced phase transitions, accompanied by the creation of 
non- zero vacuum expectation values of both elementary [32, 33] and composite bosonic 
[33- 35] or fermionic [21- 33] fields, is turned out to be important to construct the
realistic Early Universe  scenario.

It is well-known that, unfortunately, the self-consistent quantum theory of gravity does not
exist now. Because the Einstein general relativity is unrenormalizable [36- 37], we have to
find out the other models with more attractive quantum features.  
The most natural and  simplest generalization of Einstein theory is the R$^2$- gravity 
[33, 38- 44]\footnote{For the review of recent results, see [33].}.
It is renormalizable and asymptotically free theory. However, it has some disadvantages, 
such as the non-unitarity in the usual pertubative theory sense and unphysical ghosts presence. 
From the modern point of view the higher- derivative gravity is nothing but the  model including 
the next term in the low enrgy expansion of future complete quantum theory, based  perhaps, on
(super)string theory. This opinion lets us don't take into account the last problems because 
we anyway have to work below the Planck scale, naturally limiting the low energy gravity 
physics [33].

Therefore, the investigation of the DSB  both in 2D induced in the (super)string theory 
R$^2$-gravity [45- 48] and in the 4D  original version of this model provides us some 
essential information about the possible features of future complete quantum theory of gravity.
 
In the present paper we apply SDE formalism to quantum R$^2$ gravity  with the fermions
on the flat background. The ladder or rainbow approximation, when the vertex of fermion- graviton
interaction and graviton propagator are taken to be free is used. 
Both Landau- like general covariant and conformal gauges are considered in 2D case. 
The covariant gauge providing the minimal structure of graviton Green function (GF) 
is chosen for 4D spacetime. SDE are obtained and the integral equations determing 
the exact fermion GF are written evidently. The numerical analysis of their solutions is done in
details. The DSB is shown to exist in a different gauges. The dependance of dynamical fermion mass
on the coupling constant is found out. 

\section*{2. Dynamical symmetry breaking in  two dimensional R$^2$- 
gravity.}

We will consider here the theory with the following action:
$$
S= \int d^2 
x\sqrt{-g}\biggl[{\frac{1}{2M^2}}(R^2+2{\Lambda})+i\overline{\psi}
       \gamma^{\mu}(x)D_{\mu}\psi\biggr],
\eqno(1)
$$

where $R$ is the space-time curvature, $\psi$- 2D spinors, $\Lambda$- 
cosmological 
constant and the dimensional value $M$ is the fermion-- graviton 
coupling constant.

$$
D_{\mu}=\partial_{\mu}+{1\over 2}\omega^{ab}_{\mu}\sigma_{ab}
\eqno(2)
$$
is the spinor covariant derivative with the standard spin- connection 
$\omega^{ab}_{\mu}$.

Local Dirac matrices $\gamma_\mu(x)$ can be expressed through the 
usual flat ones $\gamma_{a}$ 
and tetrads $e^{\mu}_{a}(x)$ : 
$\gamma^{\mu}(x)=e^{\mu}_a(x){\gamma}^a$ and, finally, 
$\sigma_{ab}={1\over 4} [{\gamma}_a{,}{\gamma}_b]$. Greek and Latin 
indices correspond to 
the curved and flat tangent spacetimes respectively.

In accordace with the general background field method one gets:
$$
 g_{\mu\nu}={\eta}_{\mu\nu}+h_{\mu\nu}.
\eqno(3)
$$

We choose the gauge- fixing term in the folowing form [49] :

$$ S_{gf}={-\beta_1  \over {2M^2}} \int\!\! d^2 x \sqrt{-g} 
\bigl({\nabla}_{\lambda}   
h^{\lambda}_{\mu}-{\beta}_2{\nabla}_{\mu}h\bigr)\bigl({\eta}^{\mu\nu}
   {\nabla}_{\rho}{\nabla^\rho}+   
{\beta}_3{\nabla}^{\mu}{\nabla}^{\nu}\bigr)\bigl({\nabla}_{\sigma}h^{\sigma}
   _{\nu}-{\beta}_2{\nabla}_{\nu}h\bigr),
\eqno(4)
$$
where $h=h^{\mu}_{\mu}$ and  $\beta_1, \beta_2, \beta_3$ --  gauge- 
fixing dimensionless
parameters. The most useful tool for the calculation of the graviton 
Green functions
for these kinds of  complicate expressions is the projective 
operators method [50]. 
It gives

$$  G^{\mu\nu\rho\sigma}(k)=M^2
  \biggl[ \biggl(-{4\beta_1(\beta_2-1/2)^2(1+\beta_3)+1 \over \delta}  
+{4\over{(\Lambda-\beta_1k^4)k^4}}\biggr)k^{\mu}k^{\nu}k^{\rho}k^{\sigma}+
$$
$$
  +{-\beta_1 k^2 \over \Lambda(\Lambda-\beta_1 
k^4)}\biggl(\eta^{\mu\rho}k^{\nu}  
k^{\sigma}+\eta^{\mu\sigma}k^{\nu}k^{\rho}+\eta^{\nu\rho}k^{\mu}k^{\sigma}
  +\eta^{\nu\sigma}k^{\mu}k^{\rho}\biggr)+
$$
$$
  + \biggl({2 \beta_1(\beta_2-1)(\beta_2-1/2)(1+\beta_3) \over 
\delta}-{2 \over
  \Lambda k^4} 
\biggr)\biggl(\eta^{\mu\nu}k^{\rho}k^{\sigma}+\eta^{\rho\sigma}
  k^{\mu}k^{\nu}\biggr)k^2 +
$$
$$
  + \biggl({2\over \Lambda}+{\Lambda \over 
4\delta}-{\beta_1(\beta_2-1)^2
  (1+\beta_3)k^4 \over \delta }\biggr)\eta^{\mu\nu}\eta^{\rho\sigma}
- {1 \over 
\Lambda}\biggl(\eta^{\mu\rho}\eta^{\nu\sigma}+\eta^{\mu\sigma}
\eta^{\nu\rho}\biggr)\biggr],
\eqno(5)
$$
where
$$
{\delta}=-{\beta}_1 \bigl({\beta}_2-1\bigr)^2 
\bigl(1+{\beta}_3\bigr)k^8+
{{\Lambda k^4} \over 4} \biggl(1+4{\beta}_1 ({\beta}_2-1/2 )^2
({\beta}_3+1 ) \biggr).
\eqno(6)
$$
 Graviton - fermion interaction vertex has the usual form [29, 30]:

$$
{\Gamma}_{\mu\nu}(p,k)={1 \over 8 } \bigl(2p^{\lambda}+k^{\lambda} 
\bigr)
{\gamma}^{\sigma}\bigl( 2{\eta}_{\lambda\sigma}{\eta}_{\mu\nu}- 
{\eta}_{\lambda\mu}{\eta}_{\sigma\nu}- 
{\eta}_{\lambda\nu}{\eta}_{\sigma\mu}\bigr).
\eqno(7)
$$

The most general Lorentz invariant form for the inverse exact fermion 
GF is the following:
$$
S^{-1}(p)=A(p^2){\not\! p} - B(p^2),
\eqno(8)
$$
where  $A(p^2)$ and $B(p^2)$ are the unknown functions we should find 
out.

The SDE for GF (8)  in the ladder approximation are given by:

$$
S^{-1}-S_0 ^{-1}(p)=\int {d^2q \over {(2\pi)^2i}} 
{\Gamma}_{\mu\nu}(q,p-q)
S(q){\Gamma}_{\lambda\sigma}(p,q-p) G^{\mu\nu\lambda\sigma}(p-q)
\eqno(9),
$$
where $S_0(p)=1/{\not\!p}$ is the free fermion GF.

These equations determine the extremum of composite fermionic fields 
effective potential [16]:
$$
V_{eff}=-iSp\,[\,ln(S_{0}^{-1}S)-S_{0}^{-1}S+1\,]+V_2,
\eqno(10)
$$

where    
$$
V_2={1 \over 2}\int{d^2p \over (2\pi)^2}\int{d^2q \over (2\pi)^2}   
Sp\biggl[\Gamma_{\mu\nu}(p-q,q)S(q)\Gamma_{\rho\sigma}(q-p,p)S(q)\biggr]
G^{\mu\nu\rho\sigma}(p-q) 
\eqno(11)
$$
 
corresponds to the two-particle irreducable vacuum diagrams. 
Formula (11) is written in the ladder approximation, where the vertices 
and graviton GF are chosen to be free and the only fermion GF is taken to be 
exact. 

Now we have all the necessary parts of Feynman diagrams to calculate 
the final
expressions for the SDE (9) and the effective potential (10).
After the Wick rotation and tedious algebra the integral equations 
for 
structure functions $A(x)$ and $B(x)$ are obtained:

$$
A(x)=1+g\int_0^1 {A(y)dy \over {yA^2(y)+B^2(y)}}{1 \over 
x}K_{A}(x,y)
\eqno(12),
$$
$$
B(x)=g\int_0^1{B(y)dy \over {yA^2(y)+B^2(y)}} K_{B}(x,y),
\eqno(13)
$$

and the effective potential (10) is given by

$$
V_{eff}=-{T^2 \over 
4\pi}\biggl\{\int_0^1dx\biggl[\ln\biggl(A^2(x)+
{B^2(x) \over x}\biggr)-2{A(x)\biggl(A(x)-1\biggr)+B^2(x) \over
A^2(x)x+B^2(x)}\biggr]+
$$
$$
g\int_0^1{dx \over A^2(x)x+B^2(x)}\int_0^1{dy \over 
A^2(y)y+B^2(y)}
\biggl[A(x)A(y)K_A(x,y)+
$$
$$
B(x)B(y)K_B(x,y)\biggr]\biggr\},
\eqno(14)
$$
where $T$ is the ultraviolet cut off parameter,

$$
g={1 \over 64\pi}{M^2 \over T^2},\,\, x={p^2 \over T^2}, \,\,
y={q^2 \over T^2},\,\, A(p^2)=A(x), \,\, B(x)={B(p^2) \over T}.
\eqno(15)
$$

The evident form for the integral  equations kernels $K_A(x,y)$,  
$K_B(x,y)$ 
have been obtained explicitly for the arbitrary  $\beta_1$, 
$\beta_2$, $\beta_3$.
However they are too large to present them here.

 We would note only that
these kernels contain, in general, the terms with unpleasant factors 
like $(x-y)^{-1}$.
It means that the infrared divergencies caused by the virtual 
massless gravitons whose 
momentum tends to zero appear here. This type of divergences arises , 
probably, because the
quantum corrections for the interaction vertex are omitted in our 
approximation 
[21, 30].

However in the conformal and Landau- like gauges, which have the 
most 
considerable physical meaning, these divergences don't take place 
[30, 31]. 
That is why we investigate here the  $\beta_1 \to \infty$ case, 
corresponding to
Landau- like gauge. Then 
$$
K_A(x,y)=\biggl[-{\beta_2^2 \over (\beta_2-1)^2}{{xy+(x+y)
(x+y+l_2)/4}\over l_2}-{{(2\beta_2-1/2)(x-y)^2}\over 
l_1(\beta_2-1/2)}\times
$$
$$
    (x+y+l_2)-{(x-y)^2 \over l_1 l_2}\bigl((x-y)^2-l^2_2\bigr)\biggr]
    {1 \over {\sqrt{(x+y+l_2)^2-4xy}}}+
$$

$$
+\biggl[{\beta_2^2 \over {(\beta_2-1)^2}}{{xy+(x+y)(x+y-l_2)/4} 
\over l_2}
-{{(2\beta_2-1/2)(x-y)^2}\over l_1(\beta_2-1/2)} (x+y-l_2)+
$$
$$
{(x-y)^2 \over l_1 l_2}\bigl((x-y)^2-l^2_2\bigr)\biggr]
{{\tilde{\Theta}(x,y,l_2^{1/2})} \over 
{\sqrt{(x+y-l_2)^2-4xy}}}+{{2\beta_2(x+y)}
\over {l_1(\beta_2-1/2)}} \vert x-y \vert,
\eqno(16)
$$
$$
K_B(x,y)=\biggl[{\beta_2^2 \over 2(\beta_2-1)^2}\biggl(1-2{x+y 
\over l_2}
\biggr)+{(x-y)^2\over l_1(\beta_2-1/2)}\biggr]
{{\tilde{\Theta}(x,y,l_2^{1/2})} \over {\sqrt{(x+y-l_2)^2-4xy}}}+
$$
$$
\biggl[{\beta_2^2 \over 2(\beta_2-1)^2}\biggl(1+2{x+y \over l_2} 
\biggr)
+{(x-y)^2\over l_1(\beta_2-1/2)}\biggr]{1\over 
{\sqrt{(x+y+l_2)^2-4xy}}}+
$$
$$
{{-2 \vert x-y \vert }\over {l_1(\beta_2-1/2)}},
\eqno(17)
$$

where  $l_1={\Lambda \over T^4},\,\, 
l_2=\sqrt{l_1}\Biggl{\vert}{{\beta_2-1/2}
    \over {\beta_2-1}}\Biggr{\vert}$  and 

$$
\tilde{\Theta}(x,y,a)=\Theta \bigl(\,(\sqrt{x}-
\sqrt{y})^2-a^2 \bigr)-\Theta 
\bigl(a^2-(\sqrt{x}+\sqrt{y})^2\bigr).
\eqno(18)
$$

The analytical solution of thes non-linear integral equations  does 
not seem to be
possible. Therefore, we present here the results of numerical 
calculations by means
of the standard iterative procedure, described, for example, in [29- 31]  

The dependence of structural functions  $A(x)$ and $B(x)$ on the 
Euclidian momentum 
square is presented in the Fig. 1 for the different values of 
coupling constant $g$
and  fixed $l_1=4 $ and $\beta_2=1.05$. $A(x)$ (curve 1) doesn't 
almost depend 
on $g$. The only trivial solution  $B=0$ exists for function $B(x)$ 
for small $g$. 
However for $g>g_c =0.23$ the type of solution changes essentially 
and only the 
non-trivilal ones increasing with $g$ growth provide  the minimum of 
effective 
potential (14). The curves 4, 3, 2 correspond to the following values 
of 
$g= 0.25, 0.30, 0.35$  respectively. 

The dependence of dynamical mass, defined by the pole of exact  
fermion GF, 
on the gauge coupling constant $g$ is shown in the Fig. 2. It means 
that after
the analytical continuation into the pseudoeuclidian region the 
dynamical mass 
is obtained as the solution of the equation

$$
m^2 A(m^2)-B(m^2)=0.
\eqno(19)
$$

This Figure shows us clearly the typical behaviour of order parameter 
$m^2$ in the course 
of  phase transition accompanied by the creation of bifermionic 
condensate in the 
above critical region $g>g_c$.

Let us discuss now the conformal gauge, which plays the very 
important role in the
(super)string theory. In this gauge 

$$
g_{\mu\nu}=\exp( \varphi){\eta}_{\mu\nu}.
\eqno(20)
$$

Then, the graviton propagator is given by:

$$
G(k)={M^2 \over k^4+\Lambda},
\eqno(21)
$$

and the fermion- graviton interaction vertex are the following [31]:

$$
\Gamma (p,k)={1 \over 4 } \biggl( 2 {\not\! p} +{\not\! k} \biggr).
\eqno(22)
$$

The equations for the structural functions and and effective 
potential have the
same form as in the previous case (12)- (14) with:

$$
K_A(x,y)={x^2+ y^2+ 6 x y-l (x+y) \over 4 l}
{{\tilde \Theta (x,y,l^{1/2})} \over 
{\sqrt{(x+y-l)^2-4xy}}}-
$$
$$
{x^2+ y^2+ 6 x y+l (x+y) \over 4 l}
{1 \over {\sqrt{(x+y+l)^2-4xy}}},
\eqno(23)
$$
$$
K_B(x,y)=-{2 x +2y-l \over 2 l}
{{\tilde \Theta (x,y,l^{1/2})} \over 
{\sqrt{(x+y-l)^2-4xy}}}+
$$
$$
{2 x +2y+l \over 2 l} { 1 \over {\sqrt{(x+y+l)^2-4xy}}}
\eqno(24)
$$

where $l=- {\Lambda \over T^4}> 0$.

The plot of the functions $A(x)$ and $B(x)$ are presented in the Fig. 
3. The dynamical
symmetry breaking takes place in this case as well. The behaviour of 
function $A(x)$
doesn't almost depend on the value $g$ (curve 1). The non-trivial 
solutions for
function $B(x)$ appear for $g>g_c=2.0$ only. The curves 4, 3, 2 
correspond to the
 $g= 3, 4.5, 5.5 $. Figure 4 represents the dynamical mass as a 
function of coupling constant $g$.

\section*{3. Schwinger- Dyson equations in 4D higher- 
derivative gravity.}

The similar program can be realized for the four- dimensional gravity 
with the Lagrangian, containing square curvature terms ( for details, see
book [33]):
$$
S_{g}= \int d^4 x\sqrt{-g}\biggl[\alpha_1R^2+ \alpha_2 
R_{\mu\nu}R^{\mu\nu}+
\alpha_3 R_{\mu\nu\rho\sigma} R^{\mu\nu\rho\sigma}+\Lambda \biggr],
\eqno(25) 
$$
directly following [33, 49, 50]. $\alpha_1,\alpha_2,\alpha_3$ are the 
arbitrary constants here. 

We choose  the parameters of gauge fixing action:

$$ 
S_{gf}={\beta_1  \over 2} \int d^4 x \sqrt{-g} 
\bigl({\nabla}_{\lambda}   
h^{\lambda}_{\mu}-{\beta}_2{\nabla}_{\mu}h\bigr)\bigl({\eta}^{\mu\nu}
   {\nabla}_{\rho}{\nabla^\rho}+   
{\beta}_3{\nabla}^{\mu}{\nabla}^{\nu}\bigr)\bigl({\nabla}_{\sigma}h^{\sigma}
   _{\nu}-{\beta}_2{\nabla}_{\nu}h\bigr),
\eqno(26)
$$
demanding that the square on $h_{\mu\nu}$ part of action, 
$(S + S_{gf})^{(2)}$ 
won't contain the non-minimal terms. Then:
$$
(S_g+S_{gf})^{(2)}=\frac{1}{2}\int d^4x 
\sqrt{-g}h_{\mu\nu}\biggl[\frac{1}{2}
\beta_1 \beta_2 \eta^{\mu\nu} 
\eta_{\rho\sigma}\square^2-\frac{1}{2}\beta_1
\delta^{\mu\nu}_{\rho\sigma} \square ^2+
$$
$$
+\frac{\Lambda}{2}(\frac{1}{2} \eta^{\mu\nu}
\eta_{\rho\sigma}-\delta^{\mu\nu}_{\rho\sigma})\biggr] 
h^{\rho\sigma},
\eqno(27)
$$
where
$$\delta^{\mu\nu}_{\rho\sigma}=\frac{1}{2}(\delta^\mu_\rho 
\delta^\nu_\sigma
+\delta^\mu_\sigma \delta^\nu_\rho)
\eqno(28)
$$
and
$$
\beta_1=\alpha_2+4\alpha_3,\,\,\,
\beta_2=\frac{4\alpha_1+\alpha_2}{4(\alpha_1-\alpha_3)},\,\,\,
\beta_3=-\frac{2\alpha_1+\alpha_2+2\alpha_3}{\alpha_2+4\alpha_3}
\eqno(29)
$$

Then, the graviton GF is given by:
$$
G^{\mu\nu\rho\sigma}= -{1 \over \beta_1 (k^4+2\lambda )}
\biggl[ \eta^{\mu\rho} \eta^{\nu\sigma} + \eta^{\mu\sigma}
\eta^{\nu\rho} + { 2 (\beta_2 k^4 +\lambda ) \eta^{\mu\nu}
\eta^{\rho\sigma} \over k^4 (1-4 \beta_2 )-2\lambda }\biggr], 
\eqno(30)
$$

where $\lambda={\Lambda \over 2\beta_1}$. The graviton- fermion
interaction vertex has the same form as for the  covariant gauge  in 
the 2D model (7) and we obtain the same expressions for the SDE (12)- 
(13) 
and, with the accuracy up to the positive constant factor, for 
the effective potentilal (14) but, of course, different kernels:
$$
K_A(x,y)={y \over 16}+{1 \over 32 x}\biggl[ 
(x+y+l_1-\sqrt{(x+y+l_1)^2-4xy})
$$
$$
\biggl( -{l_1^2 -xy + 7(x+y)^2/4 \over 2l_1 }- {11 \over 8}(x+y) 
\biggr)+
$$
$$
+(x+y-l_1-\sqrt{(x+y-l_1)^2-4xy} \tilde
\Theta (x,y,l_1^{1/2})) \times
$$
$$
\biggl( {l_1^2 -xy + 7/4 (x+y)^2 \over 2l_1 } -
{11 \over 8} (x+y) \biggr)+
$$
$$
{9 \over 4(1-4 \beta_2)} \biggl( (x+y+l_2- 
\sqrt{(x+y+l_2)^2-4xy}) \times
$$
$$
\biggl({2xy + 1/2 (x+y)^2 \over l_2 } - {1 \over2} (x+y) \biggr)+
$$
$$ 
{9 \over 4(1-4 \beta_2)} (x+y-l_2-\sqrt{(x+y-l_2)^2-4xy}
\tilde \Theta (x,y,l_2^{1/2})) \times
$$
$$
\biggl(- {2 xy + 1/2 (x+y)^2 \over l_2 } +
{1 \over 2} (x+y) \biggr) \biggr],
\eqno(31).
$$
$$
K_B(x,y)= {9 \over 128 l_1^2 x} \biggl[ l_1 \biggl( (2x+2y -l_1)
(x+y-l_1-
$$
$$
\sqrt{(x+y-l_1)^2-4xy} \tilde \Theta (x,y,l_1^{1/2}))-
$$
$$
(2x+2y +l_1) (x+y+l_1-\sqrt{(x+y+l_1)^2-4xy}) \biggr)-
$$
$$
l_2 \biggl( (2x+2y -l_2)(x+y-l_2-\sqrt{(x+y-l_2)^2-4xy}
\tilde \Theta (x,y,l_2^{1/2}))-
$$
$$
(2x+2y +l_2) (x+y+l_2-\sqrt{(x+y+l_2)^2-4xy}) \biggr) \biggr],
\eqno(32)
$$

where $l_1^2= {2\lambda \over T^4} $ , $l_2^2= {l_1^2 \over 4\beta_2 
-1}$, $\beta_2 >1/4$, $g=(16 \pi^2\beta_1)^{-1}$.

On the Fig. 5 the plot of functions $A(x)$ (curve 1) and non-trivial 
solutions
of $B(x)$ minimized the effective potential for $g>g_c=2.5$ (curves 
2, 3, 4 
corresponds to $g= 4.5, 4.0, 3.4$) are presented. $\beta_2=2, l_1= 0.1$ 
here.
Fig. 6 represents the dymamical mass dependence on the coupling 
constant $g$.
   
\section*{4. Conclusions}
Thus, the  quantum higher-derivative gravity interacted with the fermions 
has been shown 
by the  numerical analysis to contain the phase with the dynamically 
broken chiral 
symmetry and dynamically generated  fermionic mass. The most 
important result of our 
paper is the fact that existance of such phenomena doesn't depend on 
the  gauge choice 
in the different models of R$^2$- gravity. However the values of 
$g_c$ and the 
character of phase transition are strongly depend on both the 
parameters of the model 
and the gauge type. 
\vskip 1cm
We would like to thank S. D. Odintsov for the formulation of the problem and 
helpful discussions. 

We are grateful to S. Falkenberg for the help in the preparation of the manuscript. 

Yu. I. Sh. is very much indebted to DAAD for finacial support of his visit to Leipzig 
University, where the final version of this work has been done and to Prof. B. Geyer
for hospitality. 

He  also expresses his deep gratitude to A. Letwin and R. Patov for their kind support.

\vfill\eject

\begin{figure}[h]
\centerline{\epsfig{figure=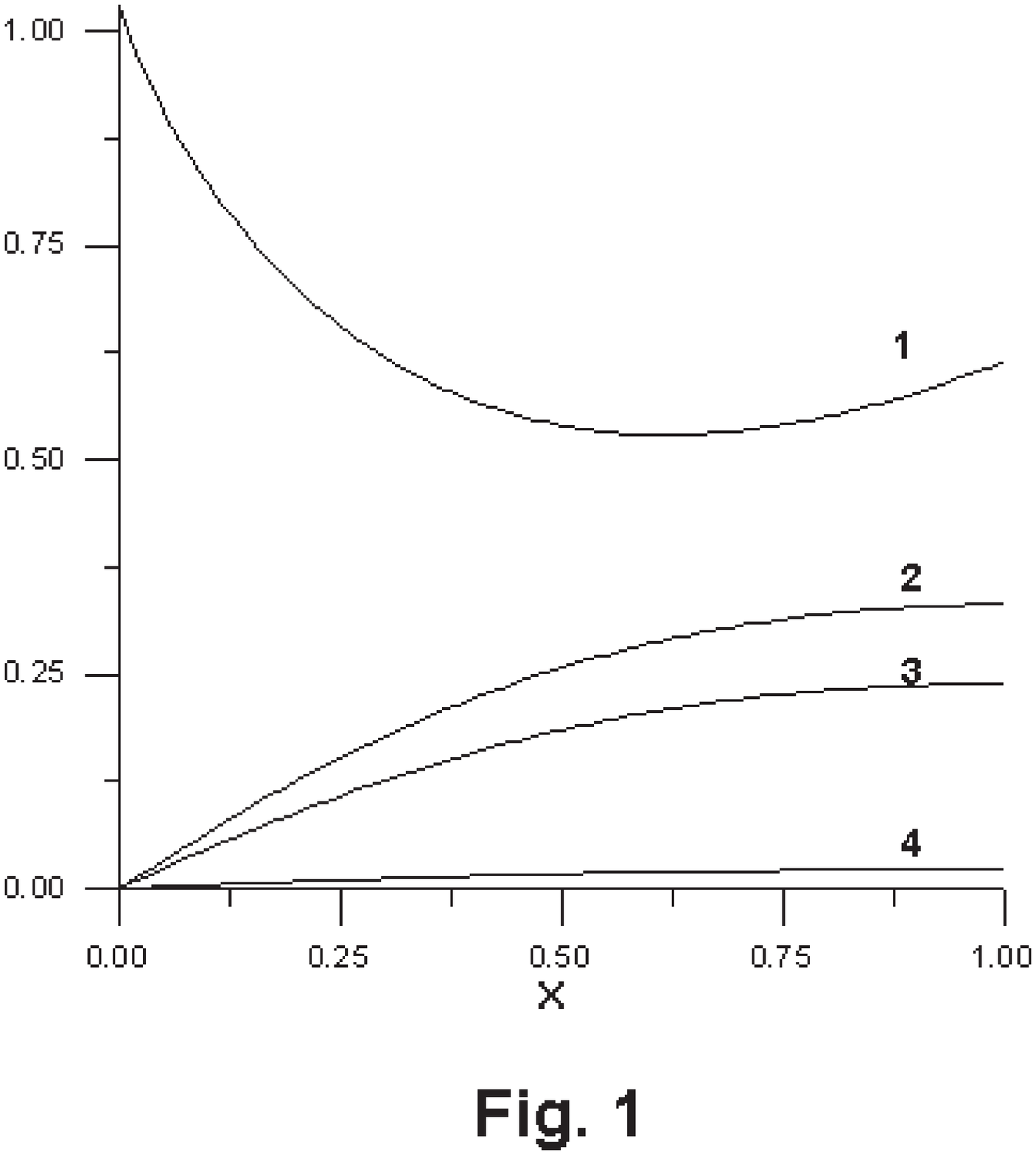, scale=.5}}
\end{figure}
\pagebreak
\begin{figure}[h]
\centerline{\epsfig{figure=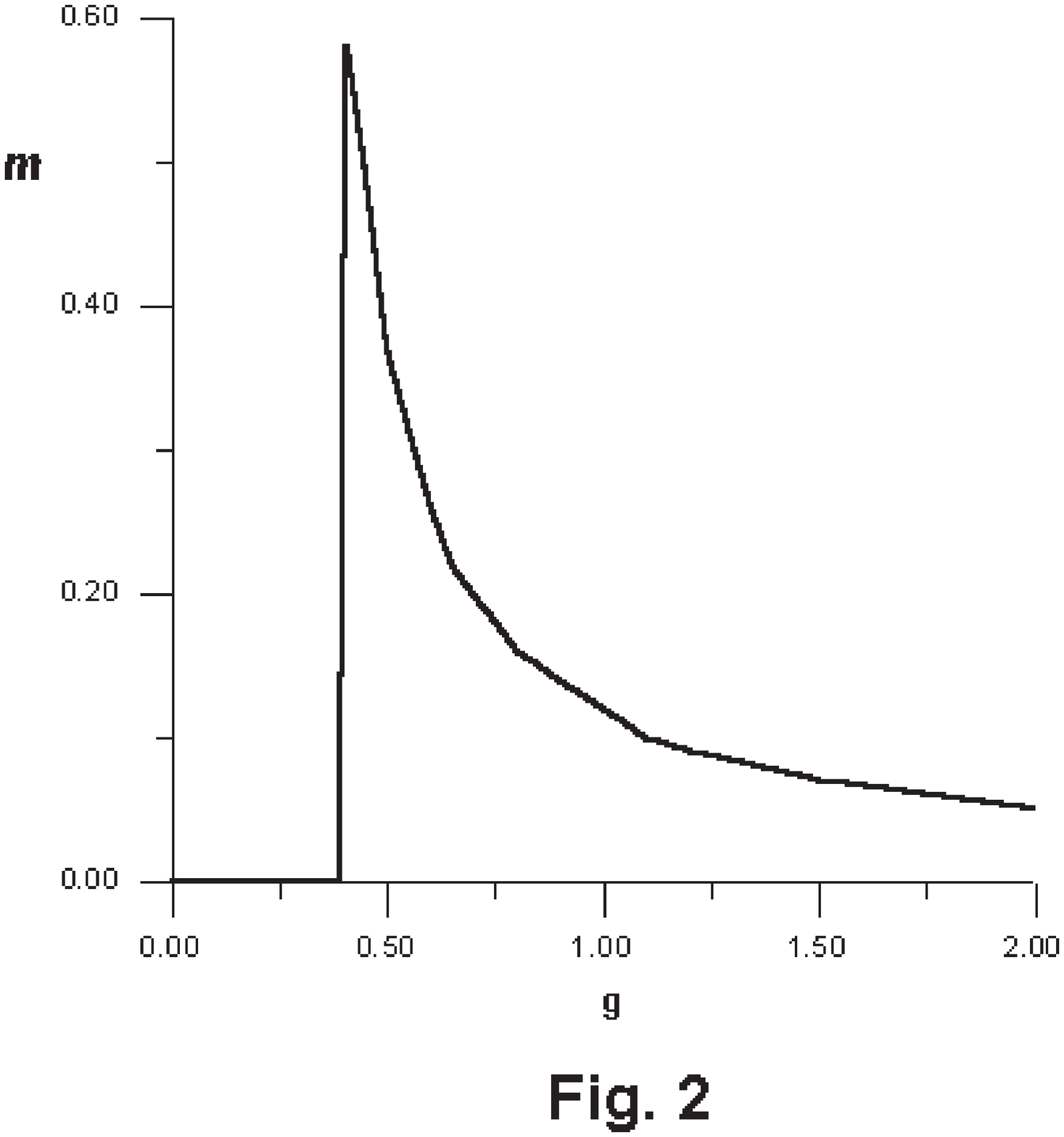, scale=.5}}
\end{figure}
\pagebreak
\begin{figure}[h]
\centerline{\epsfig{figure=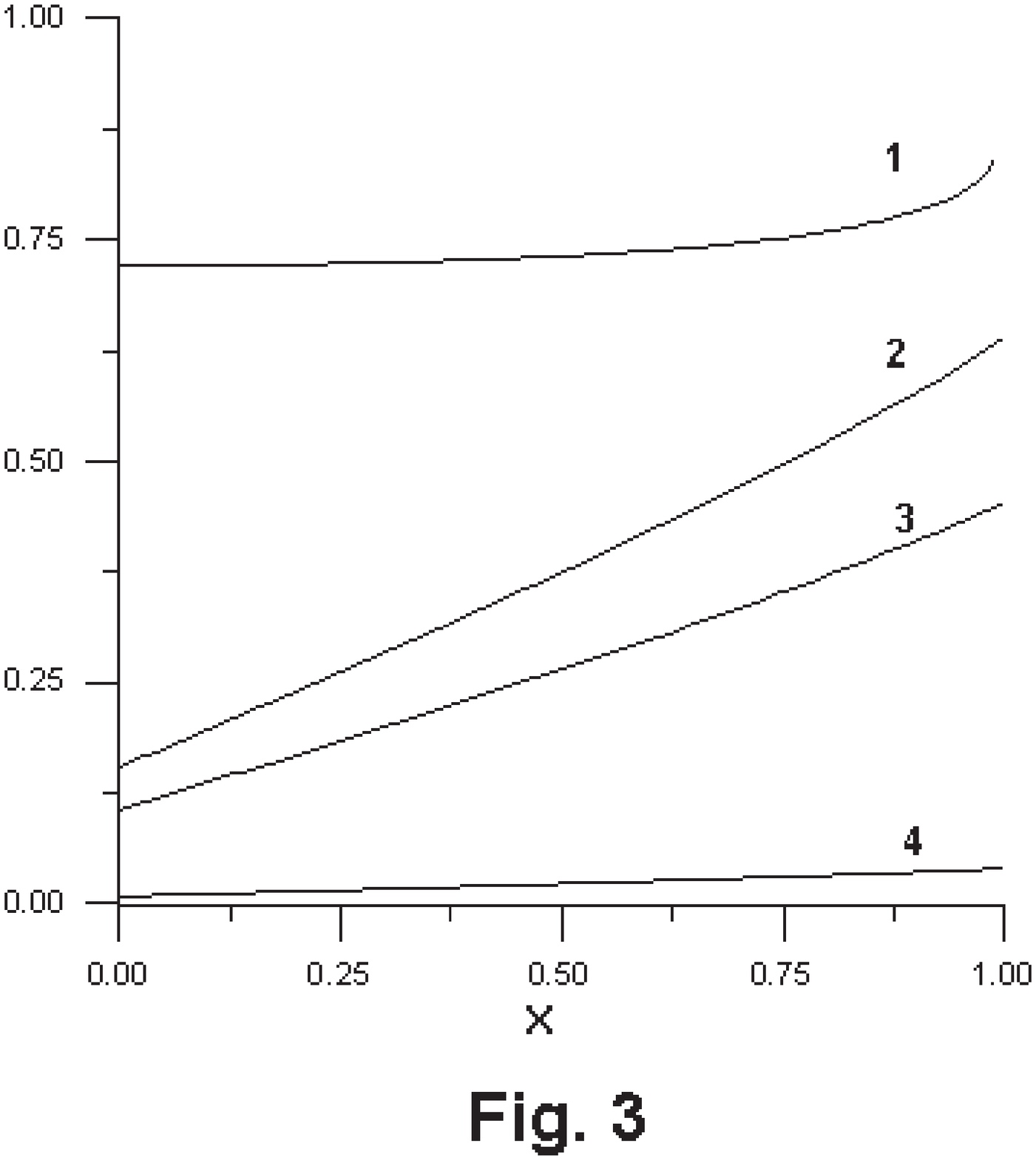, scale=.5}}
\end{figure}
\pagebreak
\begin{figure}[h]
\centerline{\epsfig{figure=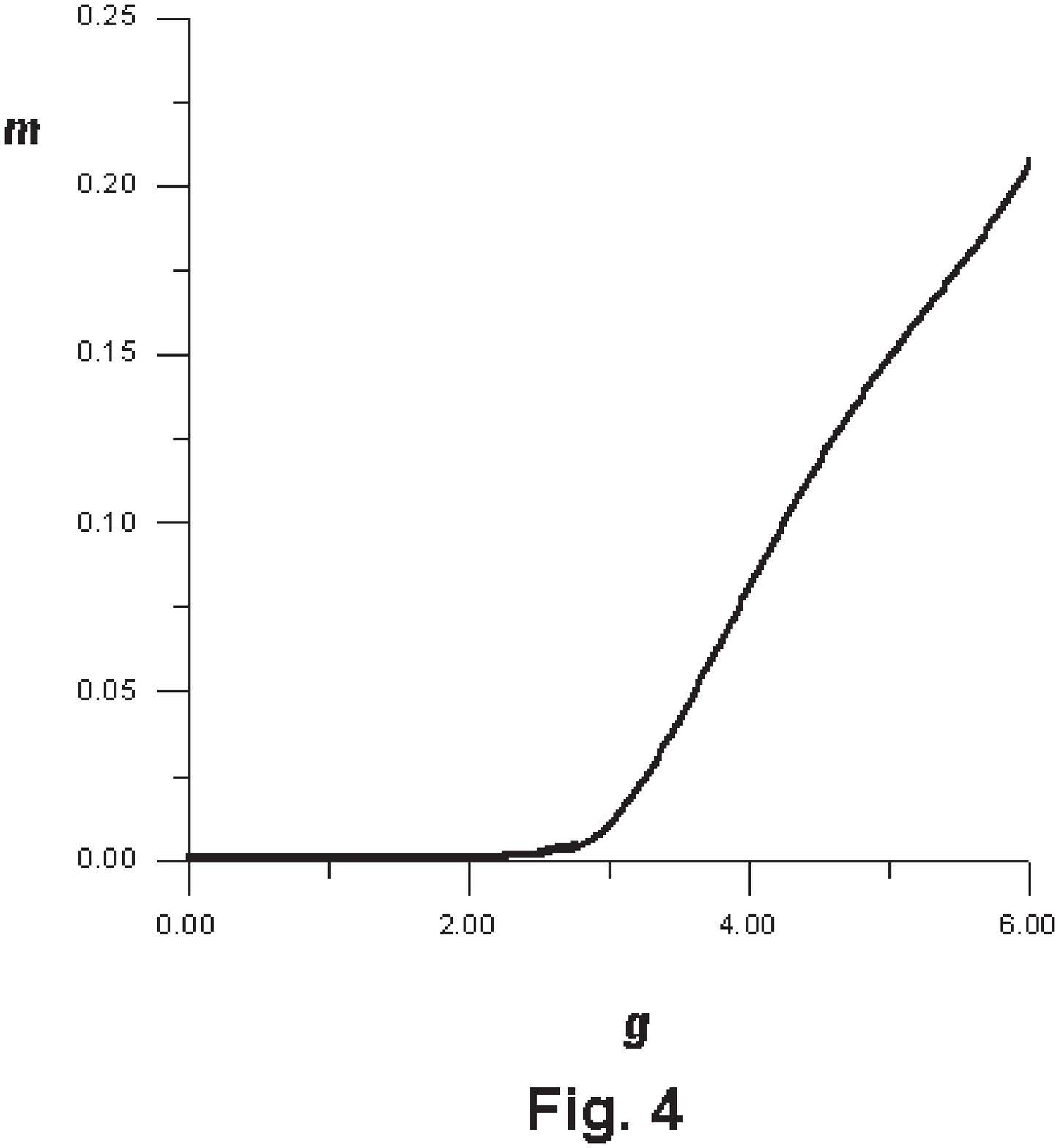, scale=.5}}
\end{figure}
\pagebreak
\begin{figure}[h]
\centerline{\epsfig{figure=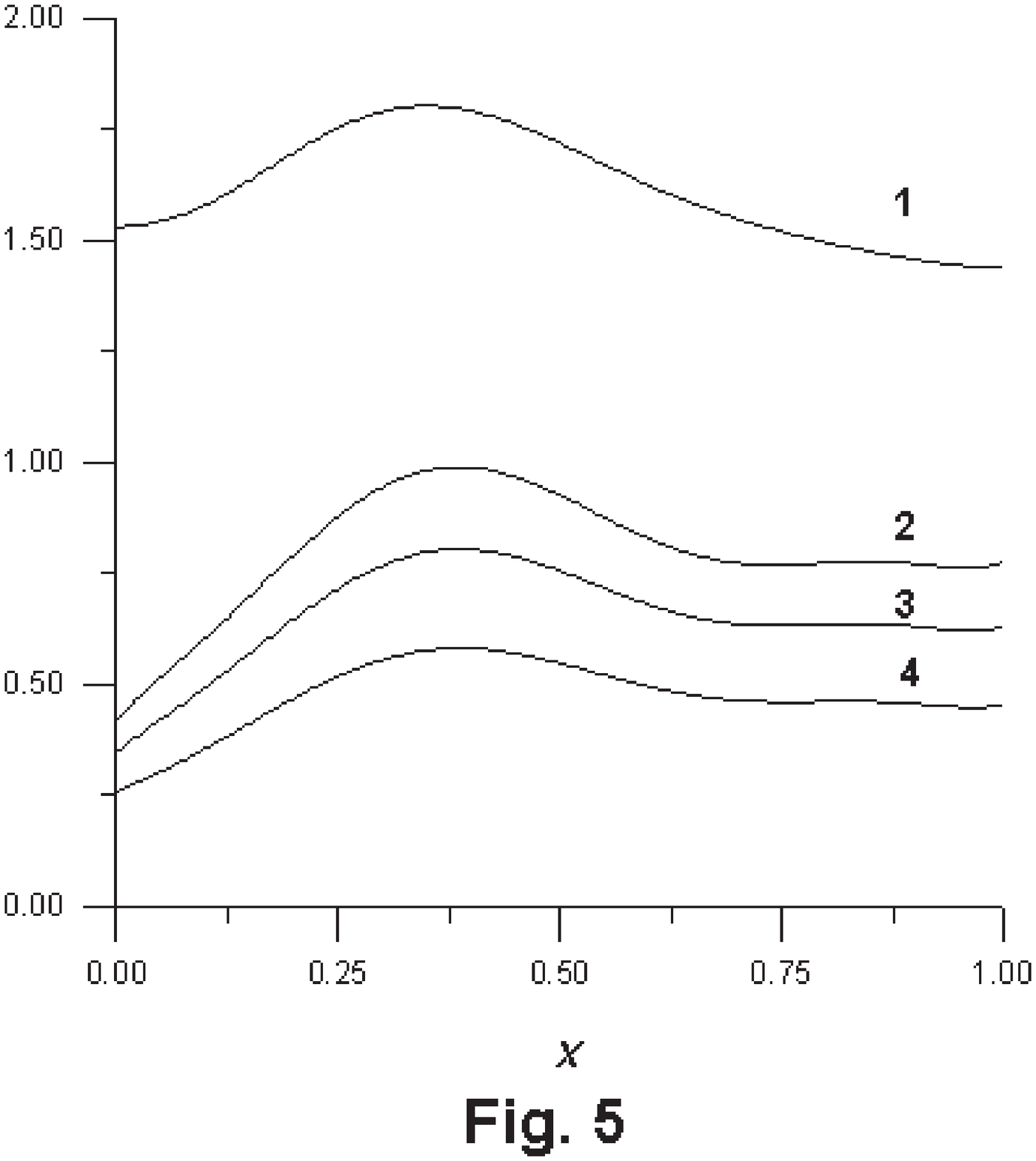, scale=.5}}
\end{figure}
\pagebreak
\begin{figure}[h]
\centerline{\epsfig{figure=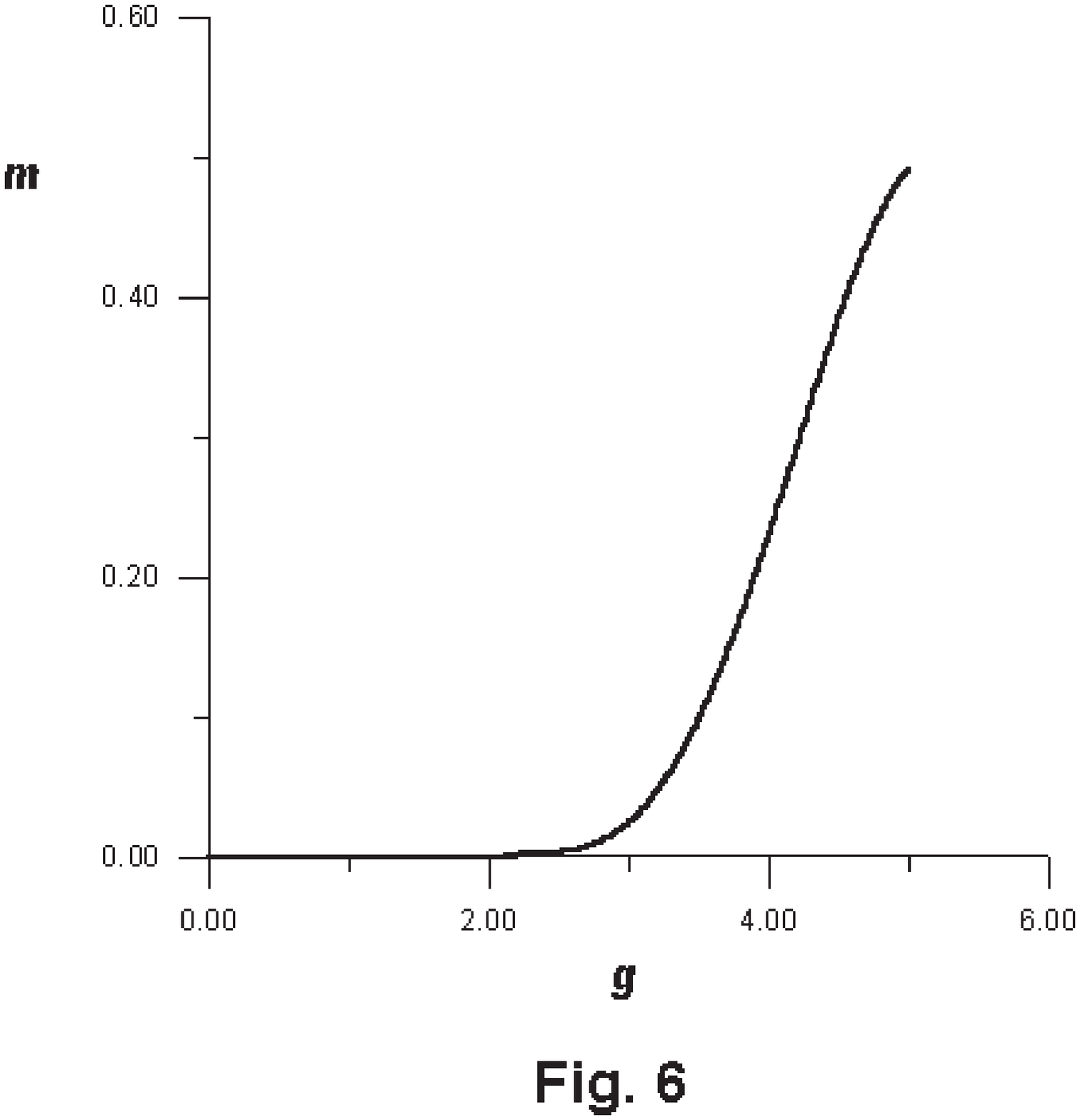, scale=.5}}
\end{figure}

\begin{thebibliography}{99}
\bibitem{1} Y. Nambu, G. Jona- Lasinio, {\it Phys. Rev.} {\bf 122}, 345 (1961). 
\bibitem{2} J. Schwinger, {\it Phys. Rev.} {\bf 125}, 2425 (1962).   
\bibitem{3} T. Maskawa, H. Nakajima, {\it Prog. Theor. Phys.} {\bf 52},1326 (1974); 
{bf 54}, 860 (1975).
\bibitem{4} R. Fukuda, T. Kugo, {\it Nucl. Phys.} {\bf B117}, 250 (1976).
\bibitem{5} P. I. Fomin, V. P. Gusynin, V. A. Miransky, Yu. A. Sitenko, 
{\it Riv. Nuov. Cim.} {\bf 6}, 1 (1983).
\bibitem{6} {\it Proceedings of the Workshop on Dynamical Symmetry 
Breaking}, ed. T. Muta, K. Yamawaki (Nagoya. 1990).
\bibitem{7} {\it Proceedings of the  International Workshop on 
Elektroweak Symmetry Breaking}, ed. W. A. Bardeen, J. Kodaira, T. Muta 
(World Scientific, Singapore. 1991).
\bibitem{8} R. W. Haymayker, {\it Riv. Nuov. Cim.} {\bf 14}, 1 (1991).
\bibitem{9} E. Fahri, R. Jackiw, {\it Dynamical symmetry breaking} 
(World Scientific, Singapore,1981).
\bibitem{10} T Muta, {\it Foundations of Quantum Chromodynamics} (World Scientific, 
Singapore, 1987).
\bibitem{11} R. D. Peccei, J. Sola, C. Wetterich, {\it Phys. Rev}. {\bf D37}, 2492 (1988).
\bibitem{12} W. A. Bardeen, C. T. Hill, M. Lindner, {\it Phys. Rev.} {\bf D41}, 1647 (1990). 
\bibitem{13} K.-I. Aoki, M. Bando, T. Kugo, K. Hasebe, H. Nakatani, {\it Prog. Theor. Phys.} 
{\bf 81}, 866(1989).
\bibitem{14} J. B. Kogut, E. Dagotto, A. Kocic, {\it Nucl. Phys.} {\bf B317}, 253; 271 (1989).
\bibitem{15} D. C. Curtis, M. R. Pennington, {\it Phys. Rev.} {\bf D48}, 4933 (1993).
\bibitem{16} J. M. Cornwall, R. Jackiw, E. Tomboulis, {\it Phys. Rev.} {\bf D10}, 2428 (1974).
\bibitem{17} J. M. Cornwall, {\it Phys. Rev}. {\bf D26}, 1453 (1982).
\bibitem{18} A. Barducci, R. Casalbuoni,  S. De Curtis, D. Dominici, R. Gatto, {\it Phys. Rev.}
{\bf D38}, 238 (1988).
\bibitem{19} S. Khebnikov, R. D. Peccei, {\it Phys. Rev.} {\bf D48}, 361 (1993).
\bibitem{20} R. P. Grigoryan, I. V. Tyutin, {\it Yad. Fiz. (Sov. J. Nucl. Phys.)} {\bf 26}, 
250 (1977).
\bibitem{21} S. D. Odintsov, Yu. I. Shil'nov, {\it Class.Quant. Grav.} {\bf 7}, 887 (1990).
\bibitem{22} T. Muta, S. D. Odintsov, {\it Mod.Phys.Lett.}. {\bf A6}, 3641 (1991).
\bibitem{23} T. Inagaki, T. Muta, S. D.  Odintsov, {\it Mod.Phys.Lett.} {\bf A8}, 2117 (1993).
\bibitem{24} E. Elizalde,  S. Leseduarte, S. D. Odintsov, {\it Phys. Rev.} {\bf D49}, 5551 (1994).
\bibitem{25} C. T. Hill, D. S. Salopek, {\it Ann. Phys.} {\bf 213}, 21 (1992).
\bibitem{26} E. Elizalde, S. D. Odintsov, Yu. I. Shil'nov, {\it Mod. Phys. Lett.} {\bf A9}, 
913 (1994).
\bibitem{27} E. Elizalde, S. Leseduarte, S. D. Odintsov, Yu. I. Shil'nov, {\it Phys. Rev.} 
{\bf D53}, 1917 (1996).
\bibitem{28} E. Elizalde, S. D. Odintsov, {\it Phys. Rev.} {\bf D51}, 5990 (1995).
\bibitem{29} O. Abe, {\it Progr.Theor.Phys.}  {\bf 73}, 1560(1985).
\bibitem{30} E. Elizalde, S. D. Odintsov, Yu. I. Shil'nov, {\it Mod. Phys.Lett.} {\bf A9}, 
2681 (1994).
\bibitem{31} E. Elizalde, S. D. Odintsov, A. Romeo, Yu. I. Shil'nov, {\it Mod.Phys.Lett.} 
{\bf A10}, 451 (1995).
\bibitem{32} A. D. Linde {\it Particle Physics and Inflationary Cosmology} (Harwood Acad., 
London, 1990).
\bibitem{33} I. L. Buchbinder, S. D. Odintsov, I. L. Shapiro, {\it Effective  action in 
quantum gravity} (IOP Publishing Ltd., Bristol, Phyladelphia, 1992).
\bibitem{34} S. W. Hawking, I. G. Moss, {\it Nucl.Phys.} {\bf B224}, 180 (1983).
\bibitem{35} I. L. Buchbinder, S. D. Odintsov, {\it Europhys.Lett.} {\bf 4}, 147 (1987).
\bibitem{36}  G.'t Hooft, M. Veltman, {\it Ann.Inst. H.Poincare} {\bf A20}, 69(1974).
\bibitem{37} S. Deser, P. van Nieuwenhuizen, {\it Phys. Rev.} {\bf D10}, 401 (1974).
\bibitem{38} K. S. Stelle, {\it Phys. Rev.} {\bf D16}, 953 (1977).
\bibitem{39} E. Tomboulis, {\it Phys. Lett.} {\bf  B70}, 361 (1977).
\bibitem{40} J. Julve, M. Tonin, {\it Nuovo Cimento}, {\bf B46}, 137 (1978).
\bibitem{41} A. Salam, J. Strathdee, {\it  Phys. Rev.} {\bf D18}, 4480 (1978).
\bibitem{42} I. Antoniadis, E. Tomboulis, {\it Phys. Rev.} {\bf D33}, 2756 (1986).
\bibitem{43} B. L. Voronov, I. V. Tyutin, {\it Yad. Fiz. (Sov. J. of Nucl. Phys.)} {\bf 39}, 
998 (1984).
\bibitem{44} I. L. Buchbinder, S. L. Lyahovich, {\it Class. Quantum Grav.} {\bf 4}, 1487 (1987).
\bibitem{45} A. M. Polyakov, {\it Phys. Lett.} {\bf B103}, 207; 211 (1981).
\bibitem{46} V. G. Knizhnik, A. M. Polyakov, A. B. Zamolodchikov, {\it Mod. Phys. Lett.}
{\bf A3}, 819 (1988).
\bibitem{47} C, G. Callan, S. B. Giddings, J. A. Harvey, A. Strominger, {\it Phys. Rev.}
{\bf D45}, 1005 (1992).
\bibitem{48} T. Muta, S. D. Odintsov, H. Sato, {\it Mod. Phys. Lett.} {\bf A7}, 3765 (1992).
\bibitem{49} V. G. Bagrov, I. L. Buchbinder, S. D. Odintsov, {\it Yad. Fiz. (Sov. J. of Nucl. 
Phys.)} {\bf 47}, 1142 (1987).
\bibitem{50} P. M. Lavrov, S. D. Odintsov, I. V. Tyutin, {\it Yad. Fiz. (Sov. J. of Nucl. Phys.)} 
{\bf 46}, 1583 (1987); S. D. Odintsov, I. N. Shevchenko, {\it Z. Phys.} {\bf C56}, 315 (1992) 
\end{thebibliography}
\end{document}